

\input harvmac.tex

\lref\S{R.P. Stanley, Adv. in Math. 77 (1989) 76.}
\lref\McDi{I.G. Macdonald, Seminaire Lotharingien, Publ. I.R.M.A.,
Strasbourg 1988.}
\lref\McDii{I.G. Macdonald, {\it Symmetric functions and Hall polynomials},
Clarendon Press (1979).}
\lref\Kadell{K.W.J. Kadell, Compos. Math. 87 (1993) 5.}

\lref\Su{B. Sutherland, J. Math. Phys. 12 (1977) 246, 251.}
\lref\Calog{F.  Calogero, J. Math. Phys. 10 (1969) 2191.}
\lref\Haldane{F.D.M. Haldane, Phys. Rev. Lett. 66 (1991) 1529.}
\lref\Forrester{P.J. Forrester, Nucl. Phys. B388 (1992) 671.}
\lref\LPS{F. Lesage, V. Pasquier and D. Serban,
``Dynamical correlation functions in the Calogero-Sutherland
model'', hep-th/9405008.}
\lref\Ha{Z.N.C. Ha, Phys. Rev. Lett. 73 (1994) 1574, cond-mat/9405063.}

\lref\FSW{P. Fendley, H. Saleur and N.P. Warner, ``Exact solution
of a massless scalar field with a relevant boundary interaction'',
to appear in Nucl. Phys. B, hep-th/9406125.}
\lref\Zamo{Al.B. Zamolodchikov, Phys. Lett. B253 (1991) 391.}
\lref\KF{C.L. Kane and M.P.A. Fisher, Phys. Rev. B46 (1992) 15233.}
\lref\WA{E. Wong and I. Affleck, Nucl. Phys. B417 (1994) 403,
cond-mat/9311040.}
\lref\IS{C. Itzykson and H. Saleur, J.Stat. Phys. 48 (1987) 449.}
\lref\AL{I. Affleck and A. Ludwig, Phys. Rev. Lett. 67 (1991) 161.}
\lref\GZ{S. Ghoshal and A.B. Zamolodchikov, Int. J. Mod.
Phys. A9 (1994) 3841, hep-th/9306002.}
\lref\M{M.L. Mehta, {\it Random matrices}, Academic Press (NY) 1991.}
\lref\KH{K.H. Kjaer and H.J.Hilhorst, J. Stat. Phys. 28 (1982) 621.}
\lref\Card{J.L. Cardy, J. Phys. A14 (1981) 1407.}

\lref\FZ{M.P.A. Fisher and W. Zwerger, Phys. Rev. B 32 (1985) 6190.}
\lref\Sc{A. Schmid, Phys. Rev. Lett. 51 (1983) 1506.}
\lref\Guin{F. Guinea, Phys. Rev. B32 (1985) 7518.}
\lref\GHM{F. Guinea, V. Hakim and A. Muramatsu,
Phys. Rev. Lett. 54 (1985) 263.}
\lref\AYH{P.W. Anderson, G. Yuval and D.R. Hamman, Phys. Rev. B1
(1970) 4464.}
\lref\MYKGF{K. Moon, H. Yi,  C.L. Kane,
S.M. Girvin and M.P.A. Fisher, Phys. Rev. Lett. 71 (1993) 4381.}
\lref\FLS{P. Fendley, A. Ludwig and H. Saleur,
``Exact Conductance through Point Contacts in the $\nu =1/3$
Fractional Quantum Hall Effect'', USC-94-12,  PUPT-94-1491,
cond-mat/9408068.}
\lref\FC{C. Callan and D. Freed, Nucl. Phys. B374 (1992) 543.}
\lref\AFL{N. Andrei, K. Furuya, and J. Lowenstein, Rev. Mod. Phys. 55
(1983) 331; \hfill\break
A.M. Tsvelick and P.B. Wiegmann, Adv. Phys. 32 (1983) 453.}
\lref\YY{C.N. Yang and C.P. Yang, J. Math.  Phys. 10 (1969) 1115.}
\lref\Ztba{Al.B. Zamolodchikov, Nucl. Phys. B342 (1991) 695.}
\lref\KM{T. Klassen and E. Melzer, Nucl. Phys. B338 (1990) 485.}

\def\t{\theta}
\def\l{\lambda}

\def\e{\epsilon}
\def\<{\langle}
\def\>{\rangle}
\def\ZZ{{\cal Z}}
\def\FF{{\cal F}}
\def\al{\alpha}

\noblackbox

\Title{\vbox{\baselineskip12pt
\hbox{USC-94-16}\hbox{SPhT-94/107}\hbox{hep-th/9409176}}}
{\vbox{\centerline{Solving 1d plasmas and 2d boundary problems}
\vskip4pt\centerline{using Jack polynomials and functional relations}}}

\centerline{P. Fendley$^1$, F. Lesage$^2$ and  H. Saleur$^{1*}$ }
\bigskip
\centerline{$^1$ Department of Physics,
University of Southern California}
\centerline{Los Angeles CA 90089-0484}
\vskip6pt
\centerline{$^2$ Service de Physique Th\'eorique de Saclay}
\centerline{F-91191 Gif sur Yvette, France}
\vskip.3in

The general one-dimensional ``log-sine'' gas is defined by
restricting the positive and negative charges of a two-dimensional
Coulomb gas to live on a circle. Depending on charge constraints,
this problem is equivalent to different
boundary field theories.

We study the electrically neutral case, which
is equivalent to a two-dimensional free boson with an
impurity cosine potential. We use two different methods:
a perturbative one based on Jack symmetric functions,
and a non-perturbative one based on the thermodynamic
Bethe ansatz and functional
relations. The first method allows us to compute explicitly
all coefficients
in the virial expansion of the free energy and
the experimentally-measurable
conductance. Some results for correlation
functions are also presented.
The second method provides in particular a surprising
fluctuation-dissipation relation
between the conductance and the free energy.

\bigskip
\bigskip\bigskip
\noindent $^*$ Packard Fellow
\Date{9/94}


\newsec{Introduction}

The general 2D classical Coulomb gas with charges restricted to live
on a circle is a recurrent problem in several areas of theoretical
physics. These include random matrix theory \M, impurity problems
(like the Kondo effect \AYH, resonant tunneling in quantum wires \KF\
and between quantum Hall edge states \refs{\MYKGF,\FLS}), and
dissipative quantum mechanics \refs{\Sc,\FZ,\GHM,\Guin,\FC}. In this
1d ``log-sine'' gas, the charges interact with a long-range interaction
proportional to the log of the sine of the separation.

Two particular cases of this Coulomb gas have been solved
analytically. The gas with only one type of charge (the Dyson gas) is
related to eigenvalue statistics for circular ensembles \M, and can be
addressed by elementary methods.  When there are two types of charges
required to alternate in space \AYH, the gas is related to the Kondo
problem \AYH, and therefore can be addressed indirectly by the Bethe
ansatz solution of the latter \AFL.

We present in this paper two methods to address the general model with
two types of charges.  The first method is direct, and uses the
recently-developed technical tool of Jack symmetric functions
\refs{\McDi,\McDii,\S,\Kadell}.  These have been used extensively in
recent works on Green functions for the Calogero-Sutherland model
\refs{\Calog,\Su,\Haldane,\Forrester,\LPS,\Ha}. The second method
 is indirect and uses the solution of the
boundary sine-Gordon theory \FSW\ via
exact $S$ matrices and the thermodynamic Bethe ansatz (TBA)
\refs{\YY,\Ztba}.

Technically, the combination of the two methods allows us to compute a
rather large number of quantities, including dynamical properties.
When they overlap, they can be compared, leading to interesting
relations between two different active areas of mathematical physics:
$1/r^2$ models and the TBA.
Physically, our solution can be applied to
a range of interesting problems, including
the experimentally-measurable resonant tunneling
between quantum Hall edge states (which was
derived using the TBA in \FLS), and the case of dissipative quantum
mechanics which will be discussed elsewhere.


We start by describing this model as a 1+1-dimensional
field theory with an impurity. Consider a Gaussian model on
an infinite cylinder with action
\eqn\act{A=-{1\over 2}\int_{-\infty}^\infty\int_0^R\ d\sigma\  d\tau\
[(\partial_\tau\phi)^2+(\partial_\sigma\phi)^2]
+2g\int_0^R \ d\tau \cos\left[\beta\phi(\sigma=0,\tau)\right].}
With periodic boundary conditions in the $\tau$-direction, this is
equivalent to a one-dimensional quantum problem at non-zero
temperature $1/R$, with an impurity at $\sigma$=0.  The
impurity coupling $g$ has a dimension, so this problem is not
conformally invariant and the interaction induces a flow from the free
boson on an infinite cylinder in the UV ($g$=0) to two decoupled free
bosons on two half-cylinders with Dirichlet boundary conditions at
their boundary in the IR ($g$ large).  A convenient quantity
describing this flow is the ``g-factor'' discussed in \AL, whose
logarithm is equal to the impurity entropy (the contribution to the
entropy which is independent of the length of the cylinder). This
``g-factor'' (which we prefer to denote $\omega$ here) is $\omega=1$
in the UV and $\omega=t^{-1/2}$ in the IR, where we define
$$t\equiv{4\pi\over \beta^2}.$$

We can study \act\ by naive perturbation theory, which exhibits the
relation with a Coulomb gas.  Defining as usual the partition function
as $Z=\int[d\phi]\ e^A$ we introduce
$${\cal Z}\equiv  {Z(g)\over Z(g=0)}.$$
Using the free field propagator on an infinite cylinder
$$\<\phi(\tau)\phi(\tau')\>=-{1\over 2\pi}
\ln\left|{\kappa R\over\pi}\sin {\pi\over R}
(\tau-\tau')\right|,$$
(where $\kappa$ is a renormalization constant) we obtain by the
standard perturbation expansion in powers of $g$
\eqn\zperti{\eqalign{{\cal Z}=
&\sum_{n=0}^\infty {g^{2n}\over  (n!)^2}
\int_0^R\ d\tau_1\ldots d\tau_n\ d\tau'_1\ldots d\tau'_n\cr
\times&{
\prod_{i<j}
\left|{\kappa R\over\pi}\sin{\pi(\tau_i-\tau_j)\over R}
\right|^{{\beta^2\over 2\pi}}
\prod_{k<l}
\left|{\kappa R\over\pi}\sin{\pi(\tau'_k-\tau'_l)\over R}
\right|^{{\beta^2\over 2\pi}}
\over
\prod_{i,k}
\left|{\kappa R\over\pi}\sin{\pi(\tau_i-\tau'_k)\over R}\right|^
{{\beta^2\over 2\pi}}
}.}}
The sum \zperti\ is the grand canonical partition function for a
classical two-dimensional Coulomb gas with {\bf two} species of
particles (with positive and negative charges) that lie on a circle
of radius $R$, and which is electrically neutral. The parameter $g$ is
the fugacity of charges, while $\beta$ controls the combination
$\hbox{charge }^2\times \hbox{inverse temperature}$.  In sect.\ 2 we
will show how these integrals can be explicitly evaluated by expanding
the integrands in terms of Jack polynomials.

We can also reformulate this as a boundary problem
(i.e.\ on  the half-cylinder) following
\WA. We introduce two new fields
$$\eqalign{\phi_e(\sigma,\tau)&={1\over\sqrt{2}}
\left[\phi(\sigma, \tau)+
\phi(-\sigma,\tau)\right]\cr
\phi_o(\sigma,\tau)&={1\over\sqrt{2}}\left[\phi(\sigma,\tau)-
\phi(-\sigma,\tau)\right],\cr}$$
so the action reads now
\eqn\newaction{\eqalign{A=-{1\over 2}
\int_{0}^\infty &\int_0^R\ d\sigma\  d\tau\
\left[(\partial_\tau\phi_e)^2+(\partial_\sigma\phi_e)^2 +
(\partial_\tau\phi_o)^2+(\partial_\sigma\phi_o)^2\right]\cr
+2g &\int_0^R \ d\tau \cos\left[{\beta\over\sqrt{2}}
\phi_e(\sigma=0,\tau)\right].\cr}}
One can of course obtain the perturbative expansion \zperti\ from
\newaction\ by using the propagator on the half-cylinder.  In the UV
both fields have Neumann boundary conditions and $\omega=\left({2\over
t}\right)^{1/2}$ while in the IR the odd field has still Neumann
boundary conditions while the even has Dirichlet boundary conditions
and $\omega={1\over\sqrt{2}}$. Notice that in the reformulation the
absolute values of $\omega$ have changed; this is presumably due to
Jacobian terms in the functional integral when redefining the new
fields. However, the ratio ${\omega_{UV}\over\omega_{IR}}=
{\omega_D\over\omega_N}=t^{-1/2}$ remains a constant.

Using this reformulation, the free energy ${\cal F}=-T\ln{\cal Z}$
can be obtained non-perturbatively using
the thermodynamic Bethe ansatz \refs{\YY,\Ztba}. The corresponding analysis
appears in
\FSW\ for $t$ integer. We will discuss this in sect.\ 4, where we will
also derive a set of functional equations which $\ZZ$ satisfies.  The
results  of TBA depend on a dimensionless variable $T/T_B$, where
$$g=\kappa' T_B^{{t-1\over t}},$$
and $\kappa'$ is an unknown renormalization constant.
The non-perturbative
free energy also contains a non-analytic term $\overline{\cal F}$
independent of $T$, which
the detailed  analysis gives as
\eqn\fbar{\overline{\cal F}={T_B\over 2\cos(\pi/2(t-1))}.}
There is also a shift because we have $\FF (g=0) =0$,
whereas the TBA is defined so that ${\cal F}_{TBA}(g\to\infty)
\longrightarrow 0$. Thus
\eqn\fftba{{\cal F}_{TBA}={\cal F}-T\ln \sqrt{t}
+\overline{\cal F}.}
This allows us to obtain the exact behavior of the
 free energy at large $g$ (large $T_B$), because the
 the power
series must be precisely balanced by $\overline{\FF}$, and thus
\eqn\largearg{{\cal Z}\approx {1\over\sqrt{t}}
\exp\left({T_B\over 2\cos({\pi/ 2(t-1)})}\right)}
in this limit. We will see in sect.\ 2 that this
behavior also follows from the expansion \zperti.

This two-dimensional local field theory problem can also be reformulated
 as a one-dimensional non-local field theory
 on a circle of circumference $R$ by integrating out the ``bulk'' degrees
 of freedom. Let us consider for instance the second (boundary) point
of view and forget about the odd field which totally decouples.
After integration the even action reads
$$A^{bdry}_e=-{\pi\over 2R^2}\int_0^R\int_0^Rd\tau d\tau'
{\phi_e(\tau)\phi_e(\tau')\over \left[\sin{\pi\over R}(\tau-\tau')\right]^2}
+2g\int_0^R d\tau\cos\left[{\beta\over \sqrt{2}}\phi_e(\tau)\right].$$
We therefore have a one-dimensional model with a sine-Gordon-type
interaction, where the Gaussian part has a $1/r^2$ interaction.
Related models have been considered in \refs{\Card, \KH}.

\newsec{Exact solution of the Coulomb gas problem using Jack polynomials}

In this section we derive all the coefficients of the
perturbative expansion of the partition function.
The result \zperti\  can be easily recast after a change of
 integration variables into
\eqn\zpertii{{\cal Z}=\sum_{n=0}^\infty x^{2n}Z_{2n},}
where  we have set
$$x=Rg
\left({\kappa R\over 2\pi}\right)
^{-{1\over t}}$$
and
\eqn\ev{\eqalign{Z_{2n}&\equiv {1\over (n!)^2}
\int_0^{2 \pi} {du_1\over 2\pi} ... {du_n\over 2\pi} {du_1'\over 2\pi}...
{du_n'\over 2\pi}
\left\vert {\prod_{i<j} 2\sin({u_i-u_j\over 2}) \prod_{k<l}
2\sin({u_k'-u_l'\over 2}) \over
\prod_{i,k} 2\sin({u_i-u_k'\over 2}) }\right\vert^{2/
t}\cr
&= {1\over (n!)^2}
\oint \prod_i \left({dz_i\over 2i\pi z_i} {dz_i'\over 2i\pi z_i'}\right)
{(\Delta(z)\overline{\Delta(z)})^{1/t}
(\Delta(z') \overline{\Delta(z')})^{1/ t}
\over \prod_{i,k} [(1-z_i \bar{z_k}') (1-z_k' \bar{z_i})]^{1
/ t} }
}
}
where $\Delta(z)$ is the  $n$-variable Vandermonde determinant
$\Delta(z)= \prod_{i<j}(z_i-z_j)$, and
$z_k=e^{i u_k}$.

To evaluate this integral, we expand the integrand in terms of
Jack polynomials \refs{\S,\McDi,\McDii}
\eqn\reljack{
\prod_{i,j} {1\over (1-r_i s_j)^a} =
\sum_\lambda b_\lambda(a) \ P_\lambda(r,a) P_\lambda(s,a) .
}
The function $P_\l(r,a)$ is a symmetric polynomial in the set of
variables $(r_1,r_2,\dots,r_n)$ which depends on a rational number
$a$.  The subscript $\l$ is a partition of an integer; this is
conveniently labeled by a Young tableau; e.g. the partition $5=2+2+1$
is labeled by the tableau with two boxes in the first row, two boxes
in the second, and one in the third.  The polynomials $P_\lambda(x,a)$
vanish if the number of parts $l(\lambda)$ of the partition $\lambda$
(i.e.\ the number of rows of the tableau) is greater than the number
$n$ of variables, so the sum runs over all partitions of all integers
with $l(\l)\le n$.  The Jack polynomials have the useful property that
their orthogonality relation involves the Vandermonde determinant:
\eqn\ortho{
\oint \prod_i {dz_i\over 2i\pi z_i} (\Delta(z) \overline{\Delta(z)})^a
\ P_\lambda(z,a) P_\mu (\bar{z},a)
= \delta_{\lambda, \mu} N_\lambda(a).
}
Hence the value of the integral follows (here $a=1/t$)~:
\eqn\resul{
Z_{2n}= {1\over (n!)^2}
\sum_{\lambda \atop l(\lambda) \leq n} b_\lambda^2 \
N_\lambda^2
}
The numerical coefficients in the foregoing expression are expressed as
a product over the boxes of the Young tableau associated with
the partition $\lambda$ \refs{\S,\McDi,\McDii}. We have
\eqn\nor{
b_\lambda^2 \  N_\lambda^2 =
c_n^2 \prod_{s\in \lambda} \left(
{j-1+{1\over t} (n-i+1)\over
j+{1\over t}(n-i) } \right)^2
}
where
$$
c_n={\Gamma({1\over t} n+1)\over
\left[\Gamma(1+{1\over t})\right]^n }$$
and  $s=(i,j)$ is the box of the tableau  at the intersection
of the $j^{\rm th}$ column and   $i^{\rm th}$ line.

One can write this product compactly using gamma functions.  Two
convenient expressions of $Z_{2n}$ follow, depending on whether one
uses partitions or their conjugates (given by interchanging the rows
and columns of the Young tableau).  One obtains
\eqn\part{
Z_{2n}= \left({c_n\over n!}\right)^2
\sum_{\lambda \atop l(\lambda)\leq n}
\prod_{i=1}^{l(\lambda)}
\left[  {\Gamma({1\over t}(n-i)+1)
\Gamma({1\over t}(n+1-i)+\lambda_i)
\over \Gamma({1\over t}(n-i+1))
\Gamma({1\over t}(n-i)+1+\lambda_i) } \right]^2
}
and alternatively, using the conjugates,
\eqn\conj{
Z_{2n}= \left({c_n\over n!}\right)^2
\sum_{\lambda \atop l(\lambda)\leq n}
\prod_{i=1}^{\lambda_1} \left( {\Gamma(n+1+t(i-1))\over
\Gamma(n+ti)}\right)^2
\left({\Gamma(n-\lambda_i'+ti)\over
\Gamma(n-\lambda_i'+1+t(i-1))}\right)^2
}
where
 $\lambda_{i+1} \leq \lambda_i$.

Consider for instance the case $n=1$~: \part\ reads then
\eqn\nun{
Z_{2}(t)=\sum_{\lambda_1=0}^\infty
\left(
{\Gamma(1/t+\lambda_1)\over \Gamma(1/t)
\Gamma(1+\lambda_1) }\right)^2
}
The sum converges only for $t>2$, so that the UV dimension of the
perturbing operator $x={\beta^2\over 4\pi}<{1\over 2}$ \IS.  This
coincides with the domain where the integrals in \zperti\ are UV
convergent. (We have no problem with IR divergences because we are on
a circle.)  The sum in \nun\ can be done explicitly; one can also
obtain its value by a direct treatment of the integral avoiding Jack
functions. The result is
\eqn\summedform{Z_2(t)=
{\Gamma(1-2/t)\over\Gamma^2(1-1/t)}.}

Let us now study the large-$n$ behavior of the series \zperti.
Using the
expression with conjugate partitions  \conj\ we find
$$
Z_{2n}= \left({c_n\over n!}\right)^2
 \sum_{\lambda \atop l(\lambda)\leq n}
\prod_{i=1}^{\lambda_1} \left(
{(n-\lambda_i'+ti-1) (n-\lambda_i'+ti-2)\ldots
(n-\lambda_i'+ti-t+1)\over
(n+ti-1)(n+ti-2)\ldots (n+ti-t+1)}\right)^2.$$
Since conjugate partitions are limited by the number $n$ we can
approximate the sum in \conj\ for large $n$ as an integral over the
variables ${\lambda_i'\over n}\equiv v_i$.  Calling the number of
boxes in the first line $\lambda_1\equiv p$ we get~:
$$
Z_{2n}\simeq \left({c_n\over n!}\right)^2 \sum_{p=0}^\infty
N^p \int_0^1 dv_1 \int_0^{v_1} dv_2 ...
\int_0^{v_{p-1}} dv_p \ \prod_{i=1}^p
\left( 1-{v_i\over (1+{ti\over N})}\right)^{2t-2}.$$
The integrand is minimized by $(1-v_i)^{2t-2}$ and maximized by $1$.
In both cases one can symmetrize over the $v_i$ to compute the
integral explicitly and one finds~:
$$ e^{n/(2t-1)} \left({c_n\over n!}\right)^2 < Z_{2n}<
e^n \left({c_n\over n!}\right)^2 .$$
Therefore, the large-$n$ behavior of the $Z_{2n}$ is fully
controlled by the $(c_n/n!)^2$ prefactor and
\eqn\asym{Z_{2n}\approx \exp\left[2n\left({1\over t}-1\right)
\log n+O(n)\right].}
An important conclusion is that for $t>2$
the radius of convergence of \zpertii\
is {\bf infinite}. Moreover approximating the sum over $n$
by an integral we find
\eqn\leading{{\cal Z}\approx
\exp\left(\hbox{constant }x^{t\over t-1}\right),}
in agreement with \largearg.  This behavior is well expected on
physical grounds. Indeed the partition function reads also $ {\cal
Z}=\exp\left({{\cal E}\over T}-{\cal S}\right) $ where $T=1/R$ is the
temperature of the equivalent one-dimensional quantum system, and
${\cal E}$ and ${\cal S}$ are the impurity energy and entropy
respectively. In the deep IR the impurity entropy converges to
$s\rightarrow\ln\omega$ and the energy, on dimensional grounds, scales
as $g^{{t\over t-1}}$.  The behavior \leading\ is the analog of the
``bulk term'' in flows between bulk critical points.

Although expressions \part\ or \conj\ are in effect a solution of the
problem, one can wonder about their practical use.
Trying to evaluate the  $Z_{2n}$ numerically, one finds that
the series converges very slowly. For example, for $t$=3, evaluating
a billion terms gives an accuracy of only about $.1\%$. Fortunately,
results are greatly improved by studying the free energy
${\cal F}=-T\ln {\cal Z}$,
whose expansion we write
\eqn\fexp{{\cal F}= T\sum_{n=0}^\infty f_{2n} x^{2n},}
The $f_{2n}$ are of course given in terms of the $Z_{2n}$.  For
example, $f_2=-Z_2$, and $f_4=-Z_4 +Z_2^2/2$. When evaluating the
$f_{2n}$ numerically for $n>1$, we find a much faster convergence. For
$t$=3, we have
\eqn\numjack{\eqalign{f_2& =-\Gamma(1/3)/\Gamma^2(2/3)
\quad\quad f_8=0.044223558\cr
f_4&=0.229454064 \qquad\qquad f_{10}=-0.022852208\cr
f_6&= -0.092261103 \qquad\quad f_{12}=0.012329254\cr}}
The $Z_{2n}$ can  be extracted from these data, and are
\eqn\numjackii{\eqalign{Z_2& =\Gamma(1/3)/\Gamma^2(2/3)
\qquad\quad Z_8=0.0618476490\cr
Z_4&=0.8378042270 \qquad\quad\ \  Z_{10}=0.01021005440\cr
Z_6&= 0.276783312 \qquad\qquad\, Z_{12}=0.00131673987\cr}}

These coefficients are enough to get a good approximation of the
properties all the way to the infrared (very large $x$) using Pade
approximants. It is then preferable to consider the entropy ${\cal
S}={\partial {\cal F}\over\partial T}$, which is bounded for $x$ large.
Keeping the coefficients through $f_{12}$, one finds for instance that
${\cal S}(x=0)-{\cal S}(x=\infty)\approx .57$, in good agreement with
the exact value $\ln\sqrt{3} \approx .549306...$

\newsec{Other quantities of interest}

The previous calculation of the partition function is the
simplest calculation which can be done using Jack symmetric
functions.
In this section we present several other calculations,
and present a conjecture for the experimentally-measurable
conductance.

\subsec{Twisted partition functions}

We have so far considered a periodic field $\phi$ on the cylinder.
We could also have winding modes such that
\eqn\winding{\phi(\sigma,\tau+R)=\phi(\sigma,\tau)+{2\pi\over\beta}p,}
where $p$ is an integer.
By splitting the field into classical and quantum parts
we obtain an action similar to \act\
but the interaction term is now
$$
2g\int_0^Rd\tau\cos
\left[\beta\phi(\sigma=0,\tau)+2\pi{p\over R}\tau\right]
$$
Defining as before ${\cal Z}(p)\equiv Z(g,p)/Z(g=0,p)$ we find
a perturbative expansion similar to \zperti\ with
however each term in the sum multiplied by
$$
\exp\left[i2\pi{p\over R}
(\tau_1+\ldots+\tau_n-\tau'_1\ldots-\tau'_n)\right].
$$
After change of variables we have the same expansion
as \zpertii\ but with
\eqn\newI{Z_{2n}(p)\equiv {1\over (n!)^2} \oint
\prod_i \left({dz_i\over 2i\pi z_i} {dz_i'\over 2i\pi z_i'}\right)
{(\Delta(z)\overline{\Delta(z)})^{1\over t}
(\Delta(z') \overline{\Delta(z')})^{1\over t}
\over \prod_{i,k} [(1-z_i \bar{z_k}') (1-z_k' \bar{z_i})]^{1
\over t}}\left({z_1\ldots z_n\over z'_1\ldots z'_n}\right)^p, }
so in the Coulomb gas language there is now  a magnetic charge
located at the center of the circle.
(We assume $p$ is positive, otherwise just
replace $p$ by $|p|$.) We now use the fact that
\eqn\newident{(z_1\ldots z_n)^p P_\lambda(z,a)=P_{\lambda+p}(z,a),}
where $\lambda+p$ means the partition $\lambda$
where $p$ columns of
length $n$ have been added. Therefore
\eqn\newIII{Z_{2n}(p)= {1\over (n!)^2}
\sum_{\lambda \atop l(\lambda) \leq n}
b_\lambda b_{\lambda+p} \
N_{\lambda}^2,}
where we use the relation $N_{\lambda}=N_{\lambda+p}$
that follows immediately
from the integral defining the norm $N$.
The coefficients $b_\lambda$ read
$$b_\lambda=\prod_{s\in \lambda}{\lambda_i-j+{1\over t}
(\lambda'_j-i+1)\over \lambda_i-j+1+{1\over t}(\lambda'_j-i)}.$$
For instance one has
\eqn\newitwo{Z_2(z)=\sum_{\lambda_1=0}^\infty
{\Gamma({1\over t}+\lambda_1) \Gamma({1\over t}+
\lambda_1+z)\over \Gamma^2({1\over t})\Gamma(1+\lambda_1)
\Gamma(1+\lambda_1+z)}={\sin{\pi\over t}\ \Gamma(1-{2\over t})\over
\sin\pi ({1\over t}+z)\Gamma(1-{1\over t}+z)\Gamma(1-{1\over t}-z)},}
for $x$ an arbitrary real number, and when $x=p$ is an integer,
\eqn\newitwobis{Z_2(p)=
{(-1)^p\Gamma(1-{2\over t})\over
\Gamma(1-{1\over t}+p)\Gamma(1-{1\over t}-p)}.}

We can also compute the partition function when electrical neutrality
is broken by some amount $Q$ (assumed integer). Defining
$${\cal Z_Q}=\lim_{\sigma\rightarrow\infty}
\sigma^{Q{\beta^2\over 2\pi}} {\int [d\phi] e^{iQ\phi(\sigma,0)}e^A
\over Z(g=0)},$$
we have
\eqn\nneutr{{\cal Z_Q}=\sum_{n=0}^\infty x^{2n+Q}
Z_{n,n+Q}.}
where $Z_{n,n+Q}$ has formally the same expression as \ev\ but there
are $n$ variables $z$ and $n+Q$ variables $z'$.  By the same
manipulations we obtain
\eqn\nneutri{Z_{n,n+Q}={1\over n! (n+Q)!}
\sum_{\lambda \atop l(\lambda) \leq n}
b_\lambda(n)b_\lambda(n+Q)N_\lambda(n)N_\lambda(n+Q).}
Observe that for $n=0$ we recover the well-known expression \M
$$
Z_{0,Q}=b_0(Q)N_0(Q)=c_Q.
$$

\subsec{Correlation functions and the conductance}

The tool of Jack polynomials should allow the perturbative evaluation
of correlation functions. However the calculation requires the
knowledge of branching coefficients which are not known yet in
general.  We will illustrate this with an example.  Consider the
two-point function of the field $\phi$ itself.
By the same perturbative approach we find
\eqn\twotcorr{\<\phi(\tau)\phi(\tau')\>_g=
\<\phi(\tau)\phi(\tau')\>_{0}
+{1\over \ZZ}{1\over 4\pi t}
\sum_{p=1}^\infty {x^{2p}\over (p!)^2}\sum_{n=1}^\infty
{(\bar{z}z')^n+(z\bar{z'})^n\over n^2}{\cal C}_{np}}
where
$${\cal C}_{np}\equiv
\oint \prod_{i=1}^p
 \left({dz_i\over 2i\pi z_i} {dz_i'\over 2i\pi z_i'}\right)
{(\Delta(z)\overline{\Delta(z)})^{1/ t}
(\Delta(z') \overline{\Delta(z')})^{1/t}
\over \prod_{i,k} [(1-z_i \bar{z_k}')
(1-z_k' \bar{z_i})]^{1/t}}
R_n(z_i, z_i')  R_n(\bar z_i',\bar z_i)
$$
$$R_n(x_i,y_i')\equiv \sum_i (x_i^n-y_i^n).$$
The calculation can be easily done for the first term $p=1$ using
\newident. One finds following the same lines as in the previous paragraph
\eqn\firstorder{
{\cal C}_{n1}=2(Z_2(n) - Z_2(0))}
For general $p$, we can still decompose the
Coulomb-interaction term between positive
and negative charges using Jack polynomials as in \reljack. However,
we cannot use the relation  \ortho\  since we have the extra
factors $R_n$. We can decompose  \refs{\Ha,\Kadell}
\eqn\polynomicrel{\sum_i z_i^n = {n\over a} \sum_{\vert \lambda \vert=n}
{\prod_{(i,j)\neq (1,1)} \left[(j-1)/a-(i-1)\right] \over
\prod_{(i,j)} \left[\lambda_j'-i+(\lambda_i-j+1)/a\right]}
P_\lambda(z,a)
.}
For our problem $a={\beta^2\over 4\pi}={1/t}$ and the
only tableaux which contribute are the ones of the form
$\lambda_1\geq\lambda_2,\ldots\geq \lambda_t,1,\ldots,1$.
The Jack polynomials in \polynomicrel\ multiply the
ones in \reljack. We therefore
 end up with the problem of determining the  coupling coefficients
$$
P_\lambda P_\mu=\sum_{\nu}g_{\lambda\mu}^\nu P_\nu.
$$
Unfortunately these coefficients are not yet  known  in general.

The two-point function of the field $\phi$ is especially
useful in physical
applications: in the 2d boundary problem the Kubo formula relates
it to the conductance \refs{\KF\FLS}, while in dissipative quantum
mechanics it is the mobility
\refs{\GHM,\Guin}. The conductance at the Matsubara
frequency $\omega_n={2\pi\over R}n$
follows from
\eqn\omegancond{G_n={2\omega_n\over t}\int_0^R
d\tau'\<\phi(\tau)\phi(\tau')\>
e^{{2i\pi\over R} n(\tau-\tau')}.}
and the $g^2$ term is easily picked up in \firstorder. The DC
conductance then is obtained by analytically continuing \omegancond\
to $n=0$, leading to
$$G={1\over t}+2{x^2\over t^2}
\hbox{ Lim }_{n\rightarrow 0}
{Z_2(n)-Z_2(0)\over n}+O(x^4).$$
Using \newitwo\
it is easy to perform the limit and one finds finally
\eqn\final{G={1\over t}-{x^2\over t^2} 2^{1-{2\over t}}
{\Gamma({1\over t})\Gamma({1\over 2})\over \Gamma({1\over 2}
+{1\over t})}}
in agreement with integral done without the Jack functions \KF.

We cannot for the moment compute the two-point function to all orders.
However, we have the following  conjecture for $G$ to all orders:
\eqn\mystery{G={1\over t}+{2x\over t^2} \left.{d^2\over dpdx}
\ln Z(g,p)\right|_{p=0},}
so the first few terms read then
$$G={1\over t}+ {2\over t^2}
\left(Z'_2 x^2+2(Z'_4-Z_2Z'_2)x^4+3(
Z'_6-Z_2'Z_4-Z_4Z'_2+Z_1'Z_2^2)x^6+\dots\right)$$
We define the continuation of $Z_{2n}(p)$ to real values of $p$
by simple substitution in \newIII, so
\eqn\limitttt{\eqalign{
 Z'_{2n}&\equiv \left.{d\over dz}Z_{2n}(z)\right|_{z=0}\cr
&={1\over (n!)^2} \sum_{\lambda \atop l(\lambda)\leq n}
b_{\lambda}^2 N_{\lambda}^2 \left[ \sum_{i=1}^n
\psi(\lambda_i+{1\over t}(n-i+1)) -
\psi(\lambda_i+1+{1\over t}(n-i))\right],\cr}}
where  $\psi(z)=\Gamma(z)'/\Gamma(z)$.
We can investigate these numbers numerically. For $t$=3,
\limitttt\ gives $Z'_2=-2.64996,
Z'_4 =-2.351,$ and $Z'_6=-0.964$.
The first agrees with \final, and
the others are in very good agreement with the conductance
calculated using the TBA in the next section.
The conjecture \mystery\ appears to be a
reasonable form of the Kubo formula \KF, but we have not
succeeded in deriving it rigorously.

To close this section we would like to remark that, although
the theory of
Jack symmetric functions generally deals with rational values
of $t$, we
expect all formulas obtained above to hold for any $t>2$ by
naive substitution.

\newsec{Non-perturbative treatment}

The boundary problem \newaction\ is integrable \GZ. One can thus find
the exact $S$ matrix for the quasiparticles of the problem \GZ, and
then use the thermodynamic Bethe ansatz \refs{\YY,\Ztba}\ to compute
the free energy \FSW\ and the conductance \FLS.  In this section, we
describe these results, and use them to derive a variety of
functional equations. These functional relations
give non-perturbative equations for the free energy, and allow one to
derive simple but  non-trivial relations among the coefficients
$Z_{2n}$. Another functional relation relates the
conductance to the free energy,
thus giving a  new fluctuation-dissipation theorem for this system.
We also find expansion coefficients at $t$=3 numerically
from the TBA as another check on our results.

The starting point of the TBA is the quasiparticle description of
a two-dimensional integrable field theory. These quasiparticles
scatter among themselves and off of the impurity with a known $S$
matrix.
At any value of $t$, the quasiparticle spectrum includes the soliton
and antisoliton, which we label by $+$ and $-$ respectively.
Moreover, at coupling $t$, there are $t-2$ ``breather'' states in the
spectrum.  The energy and momentum of these left-moving massless
particles are parametrized by rapidity variable $\t$, so
$E=-P=\mu_r\exp(-\t)$, where $\mu_+=\mu_-=\mu$ and $\mu_j=2 \mu
\sin(\pi/2(t-1))$ for the breathers.  We define the density of states
$n_r$ and the density of filled states $\rho_r$ for each quasiparticle
species $r$.  Periodic boundary conditions gives the $n_r$ as a
functional of the $\rho_r$.  The free energy can be written in terms
of these quantities; demanding it be at a minimum gives another set of
relations which determine the densities.  These relations are most
conveniently written in terms of the functions $\e_r(\t)$, which are
defined by $${1\over 1 +e^{\e_r}}\equiv {\rho_r\over n_r}.$$ Notice
that if the particles are free, $\rho_r/n_r$ is the Fermi distribution
function. However, for $t\ne 2$, the particles are not free, and the
$\e_r$ are determined by the TBA equation
\eqn\tba{\epsilon_r(\t)={\mu_r\over \mu} e^{-\t}-{1\over 2\pi}
\sum_s \int_{-\infty}^{\infty} d\t'
\varphi_{rs}(\t-\t') \ln(1+e^{-\epsilon_s(\t')}),}
where the label $s$ runs over breathers ($1\dots t-2$) and $\pm$.
The functions
$\varphi_{rs}$ for integer $t$ are given in \KM\ or \FSW.
We will not need them, because here these equations can be written in a
 much simpler form \Zamo:
\eqn\tbaii{\epsilon_r= \int_{-\infty}^{\infty} d\t'
{(t-1)\over 2\pi\cosh[(t-1)(\theta-\theta') ]}
\sum_s N_{rs}\ln(1+e^{\epsilon_s(\t')}),}
where $N_{rs}$ is the
incidence matrix of the following diagram

\bigskip
\noindent
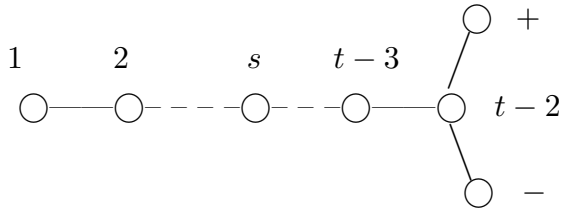
\centerline{\hbox{\rlap{\raise28pt
\hbox{$\hskip5.5cm\bigcirc\hskip.25cm+$}}
\rlap{\lower27pt\hbox{$\hskip5.4cm\bigcirc\hskip.3cm -$}}
\rlap{\raise15pt\hbox{$\hskip5.1cm\Big/$}}
\rlap{\lower14pt\hbox{$\hskip5.0cm\Big\backslash$}}
\rlap{\raise15pt\hbox{$1\hskip1cm 2\hskip1.3cm s\hskip.8cm t-3$}}
$\bigcirc$------$\bigcirc$-- -- --
--$\bigcirc$-- -- --$\bigcirc$------$\bigcirc$\hskip.3cm $t-2$ }}

\bigskip
\noindent
The dependences on the ratios $\mu_a/\mu$ seem to have disappeared
from \tbaii, but they appear as an asymptotic condition: the original
equations \tba\ indicate that the solution must satisfy
\eqn\asym{\epsilon_r \to {\mu_r\over \mu} e^{-\t}
\qquad \hbox{as}\ \t\to -\infty.}

\subsec{The partition function}

The impurity free energy is given in terms of  $\e_+$:
\eqn\freeen{{\cal F}_{TBA}={T_B\over 2\cos(\pi/2(t-1))}
-T\int {d\t\over 2\pi}
{t-1\over \cosh[(t-1)(\t-\al)]}\ln(1+e^{\epsilon_+(\t)}).}
where $\alpha\equiv\ln(T/T_B)$
\foot{This can be derived from the kernels $\kappa_a$ of
\FSW\ by using \tbaii\ along with the identity
$2\cosh y \tilde\kappa_a=\sum_b N_{ab}\tilde \kappa_b$ for
$a=1\dots t-2$ and $2\cosh y (\tilde\kappa_+ \kappa_-)=
2\tilde \kappa_{t-2} + 1$.}.
The first piece is the non-analytic term \fbar.
Since the same kernel appears in \freeen\ and \tbaii,
$\FF$ and $\ZZ$ can
be written in a simpler form for many of the $t$.
Using the relation \fftba\ between $\FF$
and $\FF_{TBA}$, we have for example
\eqn\coincZ{\eqalign{\ZZ(\al)&=\sqrt{Y_1(\al)/3} \qquad\quad t=3\cr
\ZZ(\al)&=\half(Y_2(\al))^{1/3}\quad\quad t=4\cr
\ZZ(\al)&=\sqrt{Y_3(\al)/5Y_1(\al)}\quad\quad t=5,\cr}}
where we define $Y_r\equiv \exp(\e_r)$.

We derive simple functional relations for $\ZZ(\al)$
by continuing it and the $Y_r(\al)$ into the complex
$\alpha$-plane \Zamo. Using the simple identity
\eqn\ident{\lim_{x\to 0} \left[ {\l\over
\cosh(\l\t + i\pi/2 -x)} +
{\l\over \cosh(\l\t - i\pi/2 +x)} \right] =
2\pi\delta(\t),}
we have
\eqn\funZ{\ZZ(\al+\gamma) \ZZ(\al-\gamma)
={1\over t}(1+ Y_+(\al)),}
where $\gamma\equiv i\pi/ 2(t-1)$.
Similarly, the equations \tbaii\ yield
\eqn\fnY{\eqalign{Y_+(\t+\gamma)
Y_+(\t-\gamma)&=
1+Y_{t-2}(\t)\cr
Y_{t-2}(\t+\gamma)Y_{t-2}(\t-\gamma)&=
(1+Y_{t-3}(\t))(1+Y_+(\t))^2\cr
Y_{a}(\t+\gamma)Y_a(\t-\gamma)&=
(1+Y_{a+1}(\t))(1+Y_{a-1}(\t)) \cr}}
where $a=1\dots t-3$,  we define $Y_0
\equiv 0$, and it follows from symmetry that $Y_+=Y_-$.
These equations are applicable everywhere in the complex $\t$-plane,
whereas the original TBA equations apply only in a strip
$|Im \t| < \pi/(t-1)$. The functional relations
determine the functions $Y_r$ and $\ZZ$
uniquely once the asymptotic condition \asym\ is imposed.
One can argue \Zamo\
that the functions $Y_r(\al)$
(and $\ZZ(\al)$) have the periodicity $Y_r(\al+t\gamma)
=Y_r(\al)$, which implies that they can be expanded
in powers of $(T_B/T)^{-2(t-1)/t}$ for $T_B/T$ small.
This then gives the expansions
\zpertii\ and \fexp, because $x\propto (T_B/T)^{(t-1)/t}$.

Plugging the perturbative expansions into
the functional relations \funZ\ and \fnY\ give  non-trivial
relations among the coefficients determined in sect.\ 2
by the Jack-polynomial expansions.
For $t=3,4$ these constraints can be written in a simple form
by using \coincZ. Doing a little algebra, we have
\eqn\fnZ{\eqalign{3\ZZ(\al+ i\pi/2)\ZZ(\al- i\pi/2)\ZZ(\al)
&=\ZZ(\al+ i\pi/2)+\ZZ(\al) +\ZZ(\al- i\pi/2) \qquad\quad t=3\cr
4\ZZ(\al+ i\pi/3)\ZZ(\al- i\pi/3)\ZZ(\al)
&=\ZZ(\al+ i\pi/3)+2\ZZ^2(\al) +\ZZ(\al- i\pi/3) \qquad t=4.\cr}}
These relations have the nice feature that they have lost all trace of
the quasiparticle index $r$. This is a strong hint that they can be
derived directly, without having to do the full TBA analysis. One can
also hope that there is a simple relation even for non-integer $t$
(where the TBA analysis can get quite complicated);
we make a conjecture in the next subsection.
Even though \fnZ\ are stronger relations than \funZ\ and \fnY, this
still probably isn't the end of the story: the conjecture for $t=4$
amounts to $2\ZZ(\al)\ZZ(\al+2i\pi/3)=\ZZ(\al+i\pi/3)
+\ZZ(\al-i\pi/3)$, which yields
\fnZ\ but not the other way around.

Plugging \zpertii\ into \fnZ, one finds
$$\exp\left({-3\sum_n f_{6n}x^{6n}}\right)
=\sum_n Z_{6n} x^{6n}\qquad\quad t=3.$$
This means, for example, that for $t$=3, $2Z_6=-6f_6=3Z_2Z_4- (Z_2)^3$
and $Z_{12}= -3f_{12} + 9(f_6)^2/2$. Both agree with the
Jack-polynomial expressions numerically evaluated in \numjack\ and
\numjackii.  In general, it means that that for $t$=3, the
coefficients $Z_{6n}$ are given in terms of the lower coefficients.
Similarly, for $t$=4 one finds that the coefficients $Z_{4n}$ are
determined in terms of lower coefficients. For example,
$Z_4=(Z_2)^2/3=\pi/(3\Gamma^4(3/4))$ and $9Z_8=18Z_2Z_6- Z_2^4$.  It
would certainly be interesting to have a direct proof (i.e.\ one
depending only on the expressions \part\ or \conj) of these relations.
They are certainly a hint of a much-deeper structure to the problem.

Since we cannot determine all of the coefficients from the functional
relations, a final check is to solve the TBA equations numerically and
then fit the results to a power series.  Doing a perturbative
expansion of the non-perturbative solution, one obtains
\eqn\tbaexp{{\cal F}_{TBA}=
T\sum_{n=0}^\infty k_{2n}\left({T_B\over T}\right)^{(1-{1\over t})2n}
+ \overline{\cal F}.}
We can now match these results with those of the Jack-polynomial
expansion, once comparison of the first order has determined the ratio
of the unknown constants $\kappa,\kappa'$.  We expect
\eqn\ratrel{{k_{2n}\over f_{2n}}=\left({k_2\over f_2}\right)^n.}
We evaluate the full function $f(T)$
to double-precision accuracy by solving the integral equations
\tba, plugging this into the free energy \freeen, and then fitting
this to the series \ratrel\ at large $T$.
For $t$=3, we find
\eqn\numjack{\eqalign{k_2& =-.4567084
\quad\quad\ k_8=0.000422\cr
k_4&=0.0224220 \qquad\ k_{10}=-0.00007\cr
k_6&= -0.002818 \cr}}
To the accuracy of the TBA fit, we have excellent agreement. The
scales $x$ and $T_B$ are therefore related for $t=3$  by
\eqn\scale{x^2= |k_2| {\Gamma^2(2/3)\over \Gamma(1/3)}
\left({T_B\over T}\right)^{4/3}.}

\subsec{The conductance}

In this subsection we use the non-perturbative TBA  to derive
a remarkable fluctuation-dissipation
relation of the conductance to the partition
function. This allows us to obtain  the value of
infinitely-many coefficients in the  perturbation expansion of $G$.
It also allows us to conjecture a functional relation for $\ZZ$ for
any rational value of $t$.

The TBA gives the conductance as \FLS\
\eqn\Gtba{G(\al) =\int_{-\infty}^{\infty} d\t
{t-1\over 2 \cosh^2 [(t-1)(\t-\alpha)]}
{1\over 1+ Y_+(\t)},}
where $\alpha=\ln(T/T_B)$ as before, and $Y_+$ is given
by the TBA equations \tbaii.
By using the relation
$$\lim_{x\to 0} \left[ {\l^2\over \cosh^2(\l\t + i\pi/2 -x)} -
{\l^2\over \cosh^2(\l\t - i\pi/2 +x)} \right] = -i2\pi\delta'(\t)$$
one finds
$$G(\al+\gamma)\ -\
G(\al-\gamma)=
-i{\pi\over t-1} {\del\over \del\al}{1\over 1+ Y_+(\al)},$$
where $\gamma\equiv i\pi/(2(t-1))$.
Using the relation \funZ, this gives
\eqn\GZZ{G(\al+\gamma)\ -\
G(\al-\gamma)=
-i{\pi\over t(t-1)} {\del\over \del\al}
\left[\ZZ^{-1}(\al+\gamma)\ZZ^{-1}(\al-\gamma)\right].}
This fluctuation-dissipation
relation has lost all trace of the quasiparticles of the TBA:
it is thus tempting to conjecture that it holds for all $t$,
not just the integer values
where the TBA analysis is valid.
The perturbative expansion is
$G_{pert}=\sum_n g_{2n} x^{2n}$, so we have for example
$$g_2=-Z_2{2\pi\over t^2}\cot{\pi\over t},$$
in agreement with the perturbative calculations \summedform\
and \final, which are valid for any $t>2$.

When $t\equiv p/q$ is rational, this gives many but not all of the
$g_{2n}$ in terms of the $Z_{2m}$ (with $m\le n$), because
the terms on the left-hand side of \GZZ\ vanish
when $n$ is a multiple of $p$.
For the physically-important value
$t$=3, this gives all coefficients $g_{6n+2}$ and $g_{6n+4}$;
for example
$$g_2=-{2\pi\over 9\sqrt{3}}Z_2\qquad\quad
g_4={4\pi\over 9\sqrt{3}}Z_4\qquad\quad
g_8={8\pi\over 9\sqrt{3}}(Z_8-Z_2Z_6).$$
These are in excellent agreement with a numerical
calculation of the TBA conductance.

Although the fact that some of the terms on the left-hand side of
\GZZ\ vanish for rational $t$ means we do not know how to relate these
$g_{2n}$ to the $Z_{2n}$, it does seem to imply a constraint on $Z$.
We know that $Z$ is an analytic function of $x$ for $t>2$, but we do
not know that $G$ is as well.  (One way of checking this would be to
check that the explicit perturbative expansions \limitttt\
and \part\ for $G$ and $\ZZ$ obey
the formula \GZZ.)  If $G$ is indeed analytic (so
$G$=$G_{pert}$), then it requires that these terms on the
right-hand-side also vanish, which means that $\ZZ$ should satisfy
\eqn\conject{\sum_{j=1}^p
\ZZ^{-1}(\al+2j\gamma)\ZZ^{-1}(\al+2(j-1)\gamma) =p.}
For $t$=3 this is the relation already derived in \fnZ,
but for $t=4$ it is different. Putting it together with \fnZ\ for $t=4$, we
find
the simpler relation $2\ZZ(\al)\ZZ(\al+2i\pi/3)=\ZZ(\al+i\pi/3)
+\ZZ(\al-i\pi/3)$. This relation alone implies both \conject\ and
\fnZ\ for $t=4$.
We can check the relationship \conject\ numerically.
Plugging the expansion of $\ZZ$ into \conject\ gives the coefficients
$Z_{2p}$ in terms of lower ones, which can be compared
with the expression \part. We have checked $Z_{10}$ for $t=5$ and
$t=5/2$, and find that it is indeed satisfied. This leads
us to conjecture that \conject\ is true for all $t>2$ and
rational. We also note that the relations \funZ\ and \fnY\ require
that $\ZZ$ should obey an even more
restrictive functional relationship. We have not succeeded in
finding its general form, but one can always plug the
perturbative expansions into \funZ\ and \fnY\
to derive more relations among the coefficients.

To conclude this section, we recall first that the TBA analysis is
usually made for $t$ rational only. Moreover some of the
results given above hold for $t$
integer only. However some of the functional relations we have
uncovered seem to make sense for any $t$.
Observe also that the Jack expansion and TBA  behave
differently as $t\rightarrow 2$. In the former case all integrals
just blow up, while in the latter one gets finite results for the
free energy, involving however logarithmic terms. Presumably, the TBA
gives the regularized version of the Jack computations.

\newsec{Conclusion}

Using Jack polynomials and the
thermodynamic Bethe ansatz, many properties of the
1d log-sine gas can be computed exactly, some of which are
of experimental significance. We hope that these methods can be used for
other problems with potential applications,
in particular for  dissipative
quantum mechanics. Moreover, multiple integrals
similar to those we do using
the Jack symmetric functions appear in many different kinds of
computations, so we hope that these techniques are generally applicable.

On the more formal side, it is exciting to have an example where two
different areas of mathematical physics meet.  By analogy, one might
hope that these TBA techniques can be applied to other $1/r^2$
models,
like the Calogero-Sutherland model, where Jack polynomials have been
used recently.  This overlap of techniques has led to  intriguing
relations between various quantities of Jack symmetric function
theory. For example, when $t$=4 the series \part\ can be summed to
give
$Z_4=\pi/(3\Gamma^4(3/4))$, but we have no direct proof, so we do not
know if this is a fluke or if a closed-form expression can be found
for all $t$. The overlap has also led to several simple but powerful
conjectures like \conject\ and \mystery.  One can hope that this is
evidence of a more complete mathematical structure behind the scene.

\bigskip
{\bf Acknowledgments}: H.S. is grateful to Saclay for
its warm hospitality when
this work was started, and to I. Kostov and V. Pasquier
for very useful
discussions. F.L. thanks P. Di Francesco for discussions.
This work was supported by the Packard Foundation, the
National Young Investigator program (NSF-PHY-9357207)
the DOE (DE-FG03-84ER40168), and a Canadian NSERC 67
scholarship (F.L.).

\listrefs
\end